# Crystal orbital overlap population based on all-electron *ab initio* simulation with numeric atom-centered orbitals and its application to chemical-bonding analysis in Li-intercalated layered materials

Izumi Takahara*, Kiyou Shibata, and Teruyasu Mizoguchi*

Institute of Industrial Science, The University of Tokyo, Tokyo, Japan

**Corresponding authors**

Izumi Takahara – kougen@iis.u-tokyo.ac.jp, Teruyasu Mizoguchi – teru@iis.u-tokyo.ac.jp

**Abstract**

Crystal Orbital Overlap Population (COOP) is one of the effective tools for chemical-bonding analysis, and thus it has been utilized in the materials development and characterization. In this study, we developed a code to perform the COOP-based chemical-bonding analysis based on the wavefunction obtained from a first principles all-electron calculation with numeric atom-centered orbitals. The chemical-bonding analysis using the developed code was demonstrated for $F_2$ and Si. Furthermore, we applied the method to analyze the chemical-bonding changes associated with a Li intercalation in three representative layered materials: graphite, $MoS_2$, and ZrNCl, because of their great industrial importance, particularly for the applications in battery and superconducting materials. The COOP analysis provided some insights for understanding the intercalation mechanism and the stability of the intercalated materials from a chemical-bonding viewpoint.

# 1. Introduction

Chemical-bonding among atoms composing materials is closely linked to their stability, reactivity, and electronic properties, and it has been investigated using simulations in a first principles manner. In other words, we can delve more deeply into the mechanisms underlying the emergence of materials properties by extracting information regarding chemical bonds. For the chemical-bonding analysis, some theoretical methods have been utilized, such as density of states (DOS), partial/projected DOS, and direct visualization of the wave function. One of the powerful methods in computational chemical-bonding analysis is Crystal Orbital Overlap Population (COOP) analysis [1], where orbital overlap is quantified to understand bonding and anti-bonding characteristics of chemical-bonding. Furthermore, similar concepts such as Crystal Orbital Hamiltonian Population (COHP) [2] have been devised and utilized in this context. Over the years, various DFT calculation codes have been created, whereas the development of codes for post-process COOP/COHP calculations has still been limited. The most straightforward approach to perform COOP/COHP computations is to utilizing wave function obtained from DFT calculations using localized basis sets. Computer programs for such COOP/COHP calculations have been implemented in DFT codes based on Linear Combination of Atomic Orbitals (LCAO) method like SIESTA [3,4] where atomic orbitals are employed as basis sets. Furthermore, a program named Local Orbital Basis Suite Towards Electronic-Structure Reconstruction (LOBSTER) [5–7] enables chemical-bonding analysis based on the results from plane-wave basis DFT simulations using The Vienna Ab initio Simulation Package (VASP) [8,9], Quantum Espresso [10], and ABINIT [11], by projecting plane-wave basis wave-functions onto localized basis sets. However, these methods have been developed based on non-all electron method (such as pseudopotential method) and/or delocalized basis method (such as plane-wave basis method). If we can calculate COOP based on the all-electron DFT calculation with atom-centered orbitals, a "direct" chemical-bonding analysis without any projection step including core and semi-core sates become available.

In this research, to further facilitate the chemical-bonding analysis of materials and molecules, we have newly implemented a code to calculate the COOP for arbitrary orbital pairs and arbitrary atomic pairs based on the wave-function obtained by FHI-aims [12], which is an all-electron DFT code employing numeric atom-centered orbitals as basis functions. Using the developed program, we have demonstrated the chemical-bonding analysis on simple molecular and solid-state system, $F_2$ and diamond type Si, and confirmed the validity and effectiveness.

Further, as an application of COOP analysis in materials research, we have performed a chemical-bonding analysis in Li intercalated layered materials. The phenomenon of Li intercalation into layered materials is not only utilized in battery and superconducting materials but also in modulating electrical and optical properties of materials [13–15]. It is an industrially significant phenomenon and gaining a deeper understanding of the mechanism behind Li intercalation is of great importance. In this study, we have analyzed the changes in chemical-bonding state associated with Li intercalation for representative layered materials ranging from one-elemental to ternary systems: graphite, $MoS_2$, and ZrNCl, to gain new insights into the phenomenon. We selected these systems because understanding the changes in electronic structure associated with the Li insertion is essential for developing industrially important functions in materials such as batteries and superconductors. Through the COOP based chemical-bonding analysis, we show the effectiveness of our codes.

## 2. Method of Calculations

### 2.1. Definition of COOP

The COOP, which stands for "Crystal Orbital Overlap Population", is based on the extension of the Mulliken population analysis to assign electrons to atoms, to assign electrons to chemical bonds in molecular orbital method [16,17]. COOP analysis is one of the useful and representative methods for qualitatively and quantitatively analyzing chemical-bonding in materials.

The Kohn-Sham equation, which needs to be solved self-consistently in DFT calculations, for one electron wave function $|\psi_{n,\mathbf{k}}\rangle$ in periodic system is:

$$\hat{h}^{\mathrm{KS}}|\psi_{n,\mathbf{k}}\rangle = \epsilon_n(\mathbf{k})|\psi_{n,\mathbf{k}}\rangle,$$

where the electronic Hamiltonian $\hat{h}^{\mathrm{KS}}$ is written as:

$$\hat{h}^{\mathrm{KS}} = \hat{t}_s + \hat{v}_{ext} + \hat{v}_{es} + \hat{v}_{xc}.$$

Here, $\hat{t}_s$, $\hat{v}_{ext}$, $\hat{v}_{es}$, $\hat{v}_{xc}$ are the Kohn-Sham effective single-particle kinetic energy, the external potential, the electrostatic potential of the electron density, and the exchange-correlation potential, respectively. In addition, $\mathbf{k}$, $n$, and $\epsilon_n(\mathbf{k})$ are wave vector, band index, and the corresponding eigenvalue. In non-periodic system, the Kohn-Sham equations and their solutions are not dependent of $\mathbf{k}$. When $|\varphi_{i,\mathbf{k}}\rangle$ is the basis function, the solution to the Khon-Sham equation can be written as:

$$|\psi_{n,\mathbf{k}}\rangle = \sum_i c_{i,n}(\mathbf{k})|\varphi_{i,\mathbf{k}}\rangle,$$

where $i$ is the index of basis function and $c_{i,n}(\mathbf{k})$ is the corresponding expansion coefficient. In Mulliken's approach, the overlap population $c_{i,n}(\mathbf{k})^*c_{j,n}(\mathbf{k})S_{ij,\mathbf{k}}$, which is calculated using the overlap matrix element $S_{ij,\mathbf{k}} = \langle\varphi_{i,\mathbf{k}}|\varphi_{j,\mathbf{k}}\rangle$, is defined to determine the electrons associated with the $i$-th and $j$-th orbital. The DOS weighted by the overlap population associated with $i$-th orbital is called partial density of states (PDOS):

$$\text{PDOS}_i(\epsilon) = \sum_{j,n}\int d\mathbf{k}\,\delta\big(\epsilon - \epsilon_n(\mathbf{k})\big)c_{i,n}(\mathbf{k})^*c_{j,n}(\mathbf{k})S_{ij,\mathbf{k}},$$

which enables us to obtain the contribution of $i$-th orbital to the DOS. In practical calculation, real part of the overlap population is used for the weighting. By multiplying the occupancy number and performing integration, it is possible to determine the number of electrons belonging to $i$-th orbital, which is defined as the Mulliken charge belonging to $i$-th orbital. The sum of Mulliken charge belonging to an atom corresponds to the effective charge of the atom, and the subtraction of atomic number provides the net charge of the atom.

Similarly, by considering only the pair of orbitals $i$ and $j$, and weighting the DOS with the overlap population, we can obtain:

$$\text{COOP}_{ij}(\epsilon) = \sum_{n}\int d\mathbf{k}\,\delta\big(\epsilon - \epsilon_n(\mathbf{k})\big)c_{i,n}(\mathbf{k})^*c_{j,n}(\mathbf{k})S_{ij,\mathbf{k}}.$$

The sign of COOP is related to the type of interaction between orbital $i$ and $j$, where a positive COOP corresponds to a bonding interaction and a negative COOP corresponds to an antibonding interaction. In other words, COOP diagram provides a visual representation of the bonding and anti-bonding characteristics of orbital interactions, allowing us to understand the nature of chemical-bonding in materials. The value obtained by integrating the COOP multiplied by the occupancy is called the bond overlap population (BOP), which corresponds to the number of electrons shared between orbital $i$ and $j$, and can be considered a quantity that is correlated with the strength of covalent bonds.

## 2.2. Computational details

Prior to the COOP analysis, the local structure of the materials was commonly optimized by a first-principles calculation using the plane wave basis Projector Augmented Wave (PAW) method [18], as implemented in VASP package [8,9] because some of their structures are not experimentally reported. The structural optimizations were carried out using the Generalized Gradient Approximation by Perdew, Burke, and Ernzerhof (GGA-PBE) [19] for the treatment of exchange-correlation interactions in $F_2$ molecule and diamond Si, while rev-vdW-DF2 [20–23] by Hamada was adopted to account for the van der Waals interactions in the case of graphite, $MoS_2$, ZrNCl, and their structures with Li insertion. The cutoff energy for the plane-wave expansion was set to 650 eV. For $k$-point sampling, a $k$-mesh with a spacing of 0.25 Å$^{-1}$ was used, with exception of $F_2$, for which only a single gamma point was employed. Ionic relaxation was carried out until the forces acting on each atom were reduced to less than 0.01 eV/Å.

To investigate the correlation between the chemical-bonding and intercalation behaviors, intercalation energy was calculated using the following equation:

$$E_{inter} = \frac{E[\text{LM with Li}] - E[\text{LM}] - nE[\text{Li}]}{n}.$$

Here, $E[\text{Li}]$ represents the total energy of an isolated Li atom, and $E[\text{LM}]$ represents that of a host layered material (LM), graphite, $MoS_2$, and ZrNCl in this case. $E[\text{LM with Li}]$ is the total energy of the Li-intercalated layered materials.

After performing the structural optimizations, a chemical bonding analysis was conducted using FHI-aims [12]. The electron-electron interactions were treated using GGA-PBE, and the calculation were performed at twice the *k*-point density of the structural optimization to obtain converged DOS/PDOS/COOP diagrams. After the self-consistent field simulation, the COOP (which is OP for molecule) was calculated based on the overlap population matrix elements $S_{ij,\mathbf{k}}$ ($S_{ij}$ for molecule) and expansion coefficients $c_{i,n}(\mathbf{k})$ ($c_{i,n}$ for molecule). A *minimal* basis set was employed to enhance interpretability, whereas a *light* basis set was utilized for CaC$_6$, graphite, MoS$_2$, and ZrNCl systems, allowing for analysis up to unoccupied orbitals while still maintaining interpretability.

## 3. Results and Discussion

### 3.1. F$_2$ molecule

First, to demonstrate the use of COOP in chemical-bonding analysis, we consider the example of a simple molecule, F$_2$. Since molecular system is not "crystal", we call it overlap population (OP) diagram here. Figure 1 (a) schematically illustrates the bonding and anti-bonding interactions of F$_2$ based on the simple chemical-bonding concept. $\sigma$ and $\pi$ indicate bonding interactions, while the addition of "*" denote antibonding. Based on the simple chemical-bonding concept, the interaction between $2s$ orbitals of the two F atoms give rise to bonding orbital $\sigma_2$ and anti-bonding orbital $\sigma_2^*$. In addition, the interaction between $2p_x$ orbitals result in the formation of a bonding $\sigma_3$ and anti-bonding $\sigma_3^*$ orbitals, while the interaction between $2p_y$ ($2p_z$) results in the formation of bonding $\pi$ and anti-bonding $\pi^*$ orbitals. Based on the definition of OP, we expect that the OP values to be positive for bonding orbitals such as $\sigma$ and $\pi$, and negative for anti-bonding orbitals such as $\sigma^*$ and $\pi^*$.

Figure 1 (b) shows the calculated $l$-resolved PDOS for a F atom in the $F_2$ molecule, along with the corresponding OP diagram for each orbital interaction. It can be confirmed that OP values are positive for the $\sigma$ and $\pi$ orbitals, and they are negative for the $\sigma^*$ and $\pi^*$ orbitals, which is consistent with the idea presented in Fig.1 (a). On the other hand, a detailed investigation reveals that the 2$s$-2$p$ interaction also occurs in $\sigma_2$, $\sigma_2^*$, $\sigma_3$ and $\sigma_3^*$ molecular orbitals, and those interactions are bonding, boding, antibonding, and antibonding, respectively, which cannot be explained from the simple bonding/antibonding characters of the molecular orbitals. By examining the OP diagram, we can determine the nature of molecular orbitals and understand the characteristics of chemical-bonding in detail.

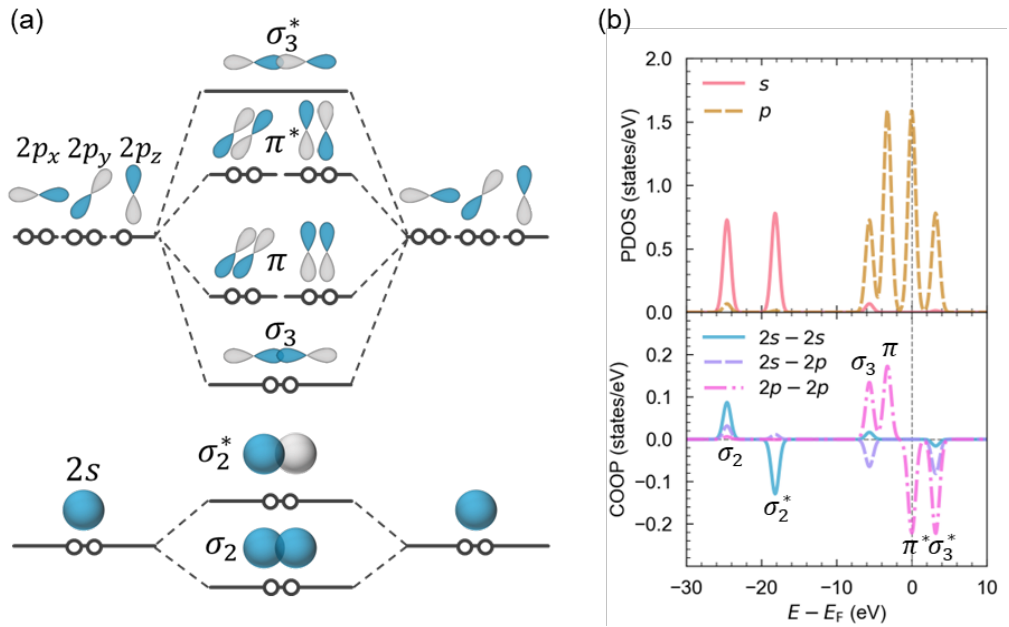

**Figure 1** (a) A schematic diagram depicting the formation of molecular orbitals from atomic orbitals in $F_2$ molecule. Sphere and dumbbell represent the $2s$ and $2p$ orbitals, respectively, and their phase was represented by blue and white colors. The white circle shows the electron. (b) Calculated $l$-resolved PDOS and COOP diagram in $F_2$. The vertical dashed line at zero represents the position of Fermi energy, $E_F$.

**3.2. Diamond type Si**

The second example is diamond-type Si, which is one of the representative covalent solid-state materials. Figure 2 shows the calculated PDOS and COOP diagrams. The results obtained using FHI-aims are presented in the leftmost column (a). To test the validity of the results, additional calculations based on the LCAO method were performed using SIESTA [3,4] and the plane-wave basis method using VASP [8,9], as shown in columns (b) and (c), respectively. SIESTA is a first-principles calculation code that utilizes numerical atomic orbitals, employing the pseudopotential method. The COOP of the identical Si crystal was obtained using a *minimal* basis set, with the $k$-grid cutoff set at 30 Å. The VASP simulation was performed with a cutoff energy of 650 eV and a $k$-mesh of $15 \times 15 \times 15$, and the PDOS/COOP was obtained through post-processing calculations using the LOBSTER [5–7].

The first row of Fig. 2 represents the $l$-resolved PDOS for a Si atom in Si crystal, and the figures on the second row shows the calculated COOP diagram regarding Si-Si interactions. In the third row, we show the COOP diagrams regarding $3s$-$3s$, $3s$-$3p$, and $3p$-$3p$ interactions. By comparing these results, it can be confirmed that the overall shapes are in good agreement across all methods. In particular, FHI-aims and SIESTA, which utilizes numeric atom-centered basis function, show very similar PDOS and COOP in both valence and conduction bands. From the COOP diagrams, we can confirm that the valence band is formed by Si-Si bonding orbitals, while the conduction band is formed by strong Si-Si anti-bonding orbitals.

Furthermore, $l$-resolved COOP can elucidate the detailed chemical-bonding characteristics of Si. In the valence band, there are three peaks, A, B, and C. Peak A, which is mainly composed of Si-3s components, is primarily attributed to the 3s-3s bonding interactions and small components of the 3s-3p bonding interactions. Peak B, which is composed of both 3s and 3p components, is attributed to the 3s-3p bonding interaction and 3s-3s antibonding interaction, whereas the valence band maximum peak C to the bonding interaction between 3p-3p orbitals.

In this way, also in solid state material, the COOP diagram can be used to visually comprehend the characteristics of the orbitals in the bands, and thereby deepen the understanding of chemical-bonding states.

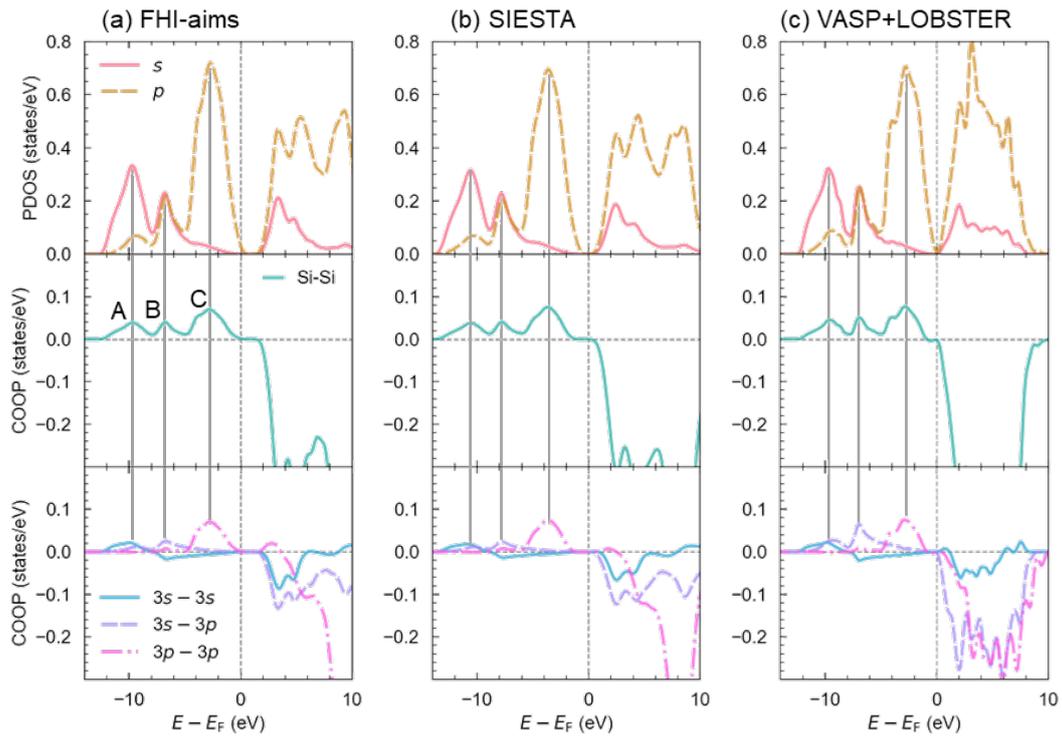

**Figure 2** (a) *l*-resolved PDOS for single Si atom, COOP diagram for Si-Si interactions, and COOP diagram illustrating the interactions between respective orbitals, calculated using FHI-aims. (b) Results from SIESTA code, and (c) Results from VASP+LOBSTER code. The vertical dashed line corresponds to the position of calculated Fermi level, and vertical solid lines show the positions of peak A, B, and C.

Although the different simulation codes show consistent PDOS/COOP in the diamond-type silicon case, the same PDOS/COOP diagrams are not necessarily guaranteed in some cases. For instance, Fig. 3 shows the results for $CaC_6$. This figure includes the total DOS, the sum of partial/projected DOS, species-specific PDOS, and COOP diagrams. Figure 3 (a) details the results computed using FHI-aims, Fig. 3 (b) outlines results from VASP+LOBSTER, and Fig. 3 (c) shows the results derived solely from VASP. Generally, when calculations are performed using a plane-wave basis set and wave functions are projected onto atoms, the sum of projected DOS does not need to identical to the total DOS, as illustrated in Fig. 3 (c). LOBSTER uses a refined projection scheme to mitigate this issue; however, when a suitable local basis is missing within the provided basis set, wave functions may be mistakenly projected onto incorrect atoms (Fig.3 (b)).

Specifically, for $CaC_6$, the first sharp peak A in the conduction band is originating from the $3d$ component of Ca atom. If these orbitals are not property included, the projection could inaccurately occur onto C atoms instead as shown in the second row of Fig. 3(b), leading to the inaccurate PDOS and COOP diagrams where anti-bonding interactions between C and Ca is not properly evaluated (the third row of Fig. 3 (b)). In our method, which employs numeric atom-centered basis function for electronic structure relaxation, it is possible to include any orbital in the electronic structure calculation, allowing the results to be directly utilized in the PDOS and COOP analysis. This capability enables the flexible chemical-bonding analysis across a wide range of materials.

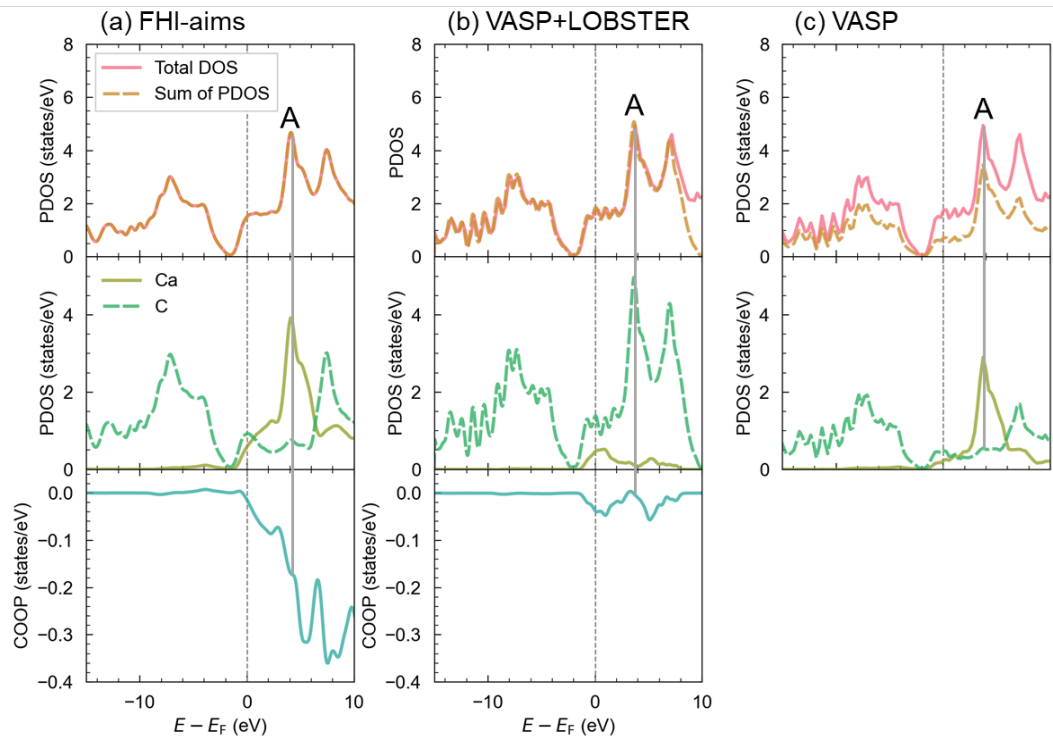

**Figure 3** (a) From top to bottom: Total DOS with sum of PDOS, atom-PDOS, and C-Ca COOP diagrams for CaC$_6$ calculated using FHI-aims, (b) VASP+LOBSTER, and (c) VASP. The vertical dashed line at zero represents the position of Fermi energy and the solid line shows the position of peak A.

### 3.3. Applications: Chemical-bonding analysis in Li-intercalated Layered Materials

As an example of the application of COOP analysis, in this study, we analyzed the changes in chemical bonding states associated with the intercalation of Li atoms in layered materials. We selected three representative layered materials, namely graphite, MoS$_2$, and ZrNCl, as shown in Fig. 3, ranging from one-elemental system to ternary systems.

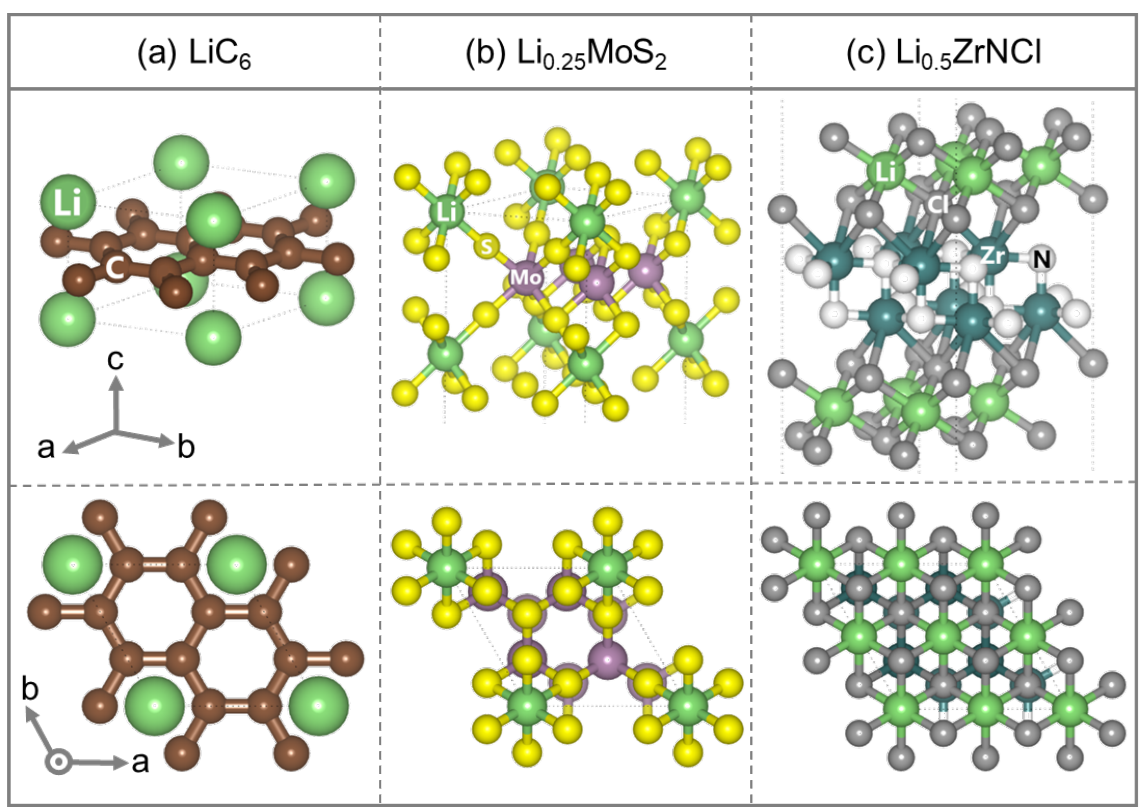

**Figure 4** Parallel projection views of (a) LiC$_6$, (b) Li$_{0.25}$MoS$_2$, and (c) Li$_{0.5}$ZrNCl visualized using VESTA[24].

### 3.3.1. Graphite

Graphite consists of stacked graphene layers held together by van der Waals interactions. Due to the relatively weak bonding between graphene layers, it easily allows the insertion of atoms or molecules into the layers[25]. The most typical application of atomic intercalation in graphite is its use as an anode material in Li-ion batteries. During the charging process, Li intercalates into the interlayer spaces of graphite, forming a graphite intercalation compound known as $LiC_6$. $LiC_6$ can stably exist, and its origin has been computationally investigated so far [26–28]. However, the specific changes in chemical bonding states associated with the intercalation process have been rarely investigated.

In this study, calculations were performed for three models with different Li concentrations: C, $Li_{0.021}C$ ($LiC_{48}$), and $Li_{0.167}C$ ($LiC_6$). We have focused only on the AA stacking structure, where both the layers that contain Li and those without Li exhibit AA-stacked structure, as it is known that the stacking structure of graphite changes from AB stacking to AA stacking upon Li intercalation. In $Li_{0.021}C$, Li atoms are intercalated every two graphene layers, which is referred to as stage-2 structure. In the case of $Li_{0.167}C$, Li atoms are intercalated between all layers as shown in Fig. 4 (a).

Figure 5 shows the calculated DOS and PDOS of graphite, $Li_{0.021}C$, and $Li_{0.167}C$. The figures on the first row shows the total DOS, and that on the second row shows the PDOS for C atom closest to Li. The figures on the third row shows the PDOS for the intercalated Li atom. By comparing the DOS and PDOS at each Li concentration, it can be observed that electrons primarily occupy the unoccupied orbitals of carbon atoms. In fact, the Mulliken charge of C atom changes from 0.000e in graphite to -0.082e in $Li_{0.167}C$, and that of Li atom changes from 0.000e in isolated Li atom to +0.493e in $LiC_6$, suggesting the emergence of the ionic bonding between Li and C atoms.

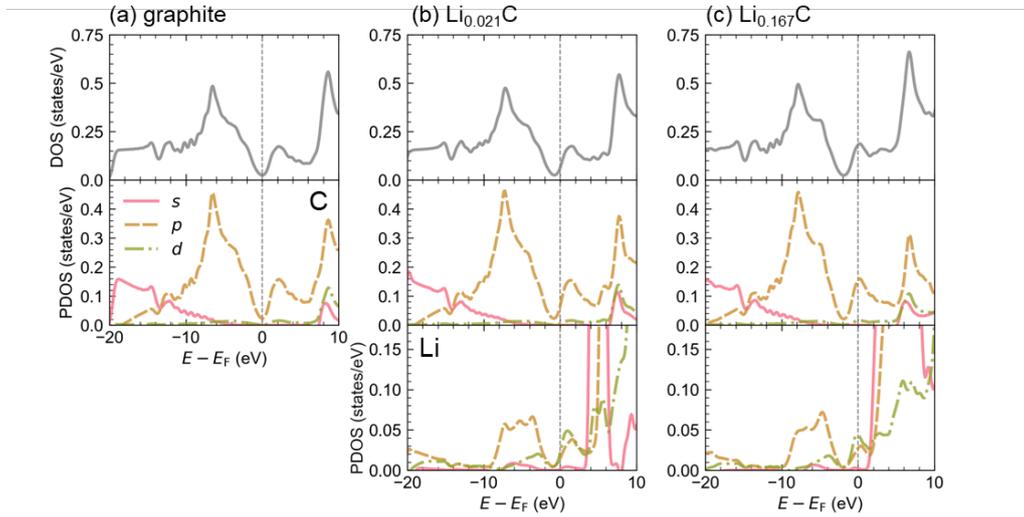

**Figure 5** Calculated total DOS and atom-PDOS in (a) graphite, (b) $Li_{0.021}C$, and (c) $Li_{0.167}C$. The vertical dashed line corresponds to the position of Fermi energy ($E_F$). The total DOS has been scaled to be per C atom.

Figure 6 shows the results of COOP analysis at each Li concentration. The first, second, and third rows of the figure correspond to the total DOS, COOP diagram for the C-C interactions adjacent to Li, and COOP diagram for the Li-C interactions, respectively. From Fig. 6, it is observed that the absolute value of COOP between Li and C is strengthened upon increasing Li concentration at the energy indicated by "p1", indicating that the hybridization occurs between orbitals of Li and C. Additionally, since its COOP is relatively positive, the interaction between Li and C is of a bonding nature. Previous study has suggested that the stable existence of $LiC_6$ is due to the formation of covalent bonds between Li and C, with charge density maps serving as the evidence for this argument [26]. In this study, the COOP analysis successfully quantitated the bonding interactions between Li and C, and revealed that the covalent bonds gradually strengthen as the concentration of Li increases. On the other hand, the COOP diagram also reveals that the electrons start to occupy the C-C anti-bonding orbitals (second row of Fig.6), which likely weaken the covalent bonds between them. In fact, the BOP values of C-C interaction decrease from 0.500e in graphite to 0.460e in $Li_{0.167}C$. The intercalation energy increases from -2.213 eV in $Li_{0.021}C$ to -1.887 eV in $Li_{0.167}C$, and the increase in Li content contributes to destabilization, which is possibly correlated with the destabilization of C-C bonds in terms covalent bonding.

To summarize, from the above results, it is considered that when Li is intercalated in graphite, the charge is transferred from Li to C from the viewpoint of Mulliken charge, and an ionic interaction occurs. In addition, COOP analysis revealed that the intercalation of Li induces covalent bonding interactions between Li and C, while anti-bonding interactions occur between C atoms, resulting in the destabilization of the structure.

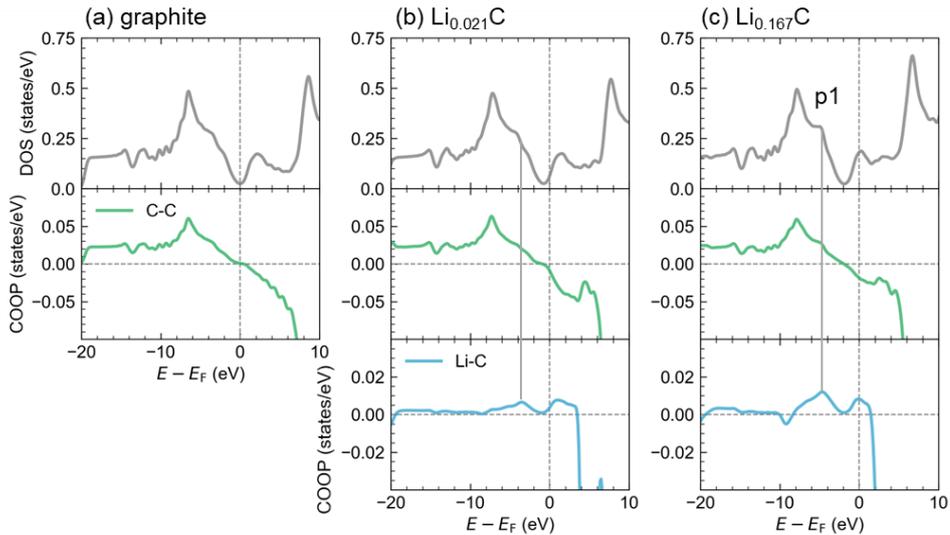

**Figure 6** Calculated total DOS and COOP curves in (a) graphite, (b) $Li_{0.021}C$ and (c) $Li_{0.167}C$. The vertical dashed line corresponds to the position of Fermi energy ($E_F$). The total DOS has been scaled to be per C atom.

### 3.3.2. 2H-MoS$_2$

MoS$_2$ is a representative two-dimensional compound of binary systems and belongs to the class of transition metal dichalcogenides (TMDs). Similar to graphite, Li can be inserted between the layers of MoS$_2$, and the phenomenon of Li intercalation is utilized for the applications in battery materials, modulation of electrical properties, and exfoliation of MoS$_2$ layers[13]. It is known that MoS$_2$ undergoes successive structural phase transition from 2H to 1T, and then to 1T' phase [29]. In this research, we placed emphasis on the initial phase, 2H phase, to investigate the interaction between Li and the MoS$_2$ layer in detail. Specifically, we focused on the region with relatively low Li concentrations: MoS$_2$, Li$_{0.056}$MoS$_2$, and Li$_{0.250}$MoS$_2$, and examined the changes in the chemical-bonding states associated with the Li interactions. In Li$_{0.056}$MoS$_2$, Li is inserted every two layers, while in Li$_{0.250}$MoS$_2$, which has the highest Li concentration, Li is inserted every interlayer, as shown in Fig. 4 (b).

Figure 7 shows the calculated total DOS and PDOS for each composition. The first, second, and third rows of the figure corresponds to the total DOS, PDOS of Mo and S, and PDOS of Li, respectively. Unlike in the case of graphite, significant changes in the shape of DOS and PDOS curves are not observed with the change in the Li concentration in MoS$_2$. On the other hand, we observed a consistent change in Mulliken charges. The Mulliken charge of Li increases from +0.175e to +0.216e with the increase in the Li concentration, acquiring a more positive charge. Besides, the Mulliken charge of Mo remains unchanged, while the Mulliken charge of S exhibits a monotonic decrease from -0.144e in MoS$_2$ to -0.192e Li$_{0.25}$MoS$_2$. This suggests that the electron provided by the Li intercalation is mainly transferred to the neighboring S atoms within the range of Li concentrations considered in this study.

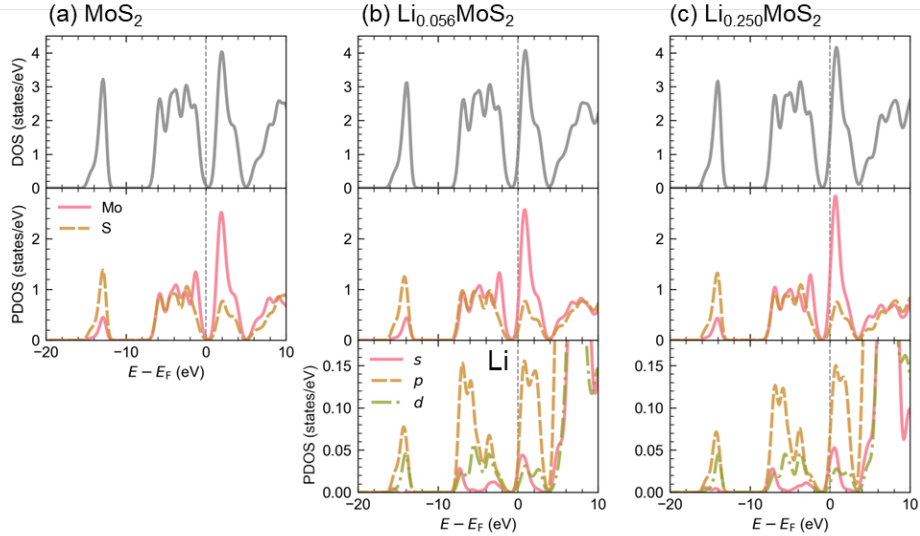

**Figure 7** Calculated total DOS and atom-PDOS for Mo, S, and Li in (a) $MoS_2$, (b) $Li_{0.056}MoS_2$, and (c) $Li_{0.250}MoS_2$. The vertical dashed line corresponds to the position of Fermi energy ($E_F$). The total DOS has been scaled to be per $MoS_2$.

Figure 8 shows the results of COOP analysis. Total DOS, COOP diagram about Mo-S interactions, and that about Li-S interactions for each Li concentration are presented. From Fig. 8, it can be observed that there is generally a bonding interaction between Li and S, and the extent of this interaction does not change even with the variations in the Li concentration. Furthermore, the Mo-S COOP features remain unchanged, indicating that interaction of Li does not affect the chemical-bonding state between Mo and S. In fact, the BOP between Mo and S remains almost unchanged. Similarly, the change in intercalation energy was also found to be small. Even with the change in Li concentration from $Li_{0.056}MoS_2$ to $Li_{0.25}MoS_2$, the intercalation energy increased by only 0.020 eV.

In the case of $MoS_2$, it was observed that Li insertion led to the transfer of electrons from Li to S.

However, even with an increase in Li concentration, the COOP diagrams for both Li-S and Mo-S interactions were not significantly affected. Furthermore, the change in the intercalation energies was also small. This indicates that the changes in the chemical bonding state are closely related to the stability of the intercalated structures.

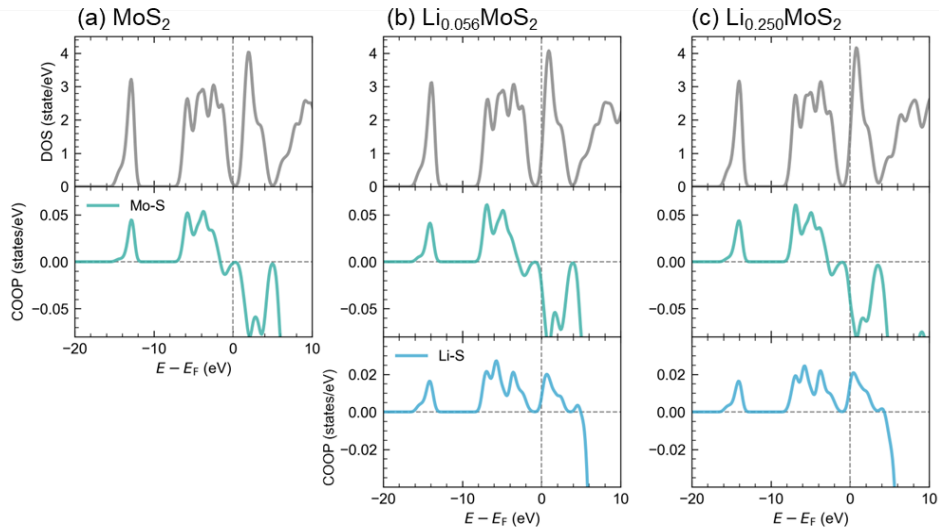

**Figure 8** Calculated total DOS and COOP diagrams about Mo-S interactions and Li-S interactions in (a) $MoS_2$, (b) $Li_{0.056}MoS_2$, and (c) $Li_{0.250}MoS_2$. The vertical dashed line corresponds to the position of Fermi energy ($E_F$). The total DOS has been scaled to be per $MoS_2$.

### 3.3.3. ZrNCl

Finally, we have applied the COOP analysis for a ternary layered material, ZrNCl, which can also allow Li intercalation between the ZrNCl layers, leading to the emergence of superconductivity[30,31]. Because it is known that stacking structure changes from SmSI type to YOF type upon Li intercalation[32], we considered YOF type structure with three different Li concentrations: ZrNCl, $Li_{0.042}ZrNCl$, and $Li_{0.500}ZrNCl$. In $Li_{0.042}ZrNCl$, Li is inserted every three layers, while in $Li_{0.500}ZrNCl$,

Li exists in every interlayer as shown in Fig. 4 (c).

Calculated total DOS and atom PDOS are shown in Fig. 9. The figures on the first row represents the total DOS, and the second and third rows represent the PDOS of Zr and Cl bonded to Li, respectively. In the case of $Li_{0.042}ZrNCl$, the dilute in-plane configuration of Li results in several symmetrically distinct atom sites for Zr, Cl, and N, and we selectively investigated PDOS of atoms close to intercalated Li. From PDOS in Fig. 8, it can be observed that the electronic states of all elements undergo changes with an increasing Li content, and changes in the peaks occur at the same energy levels in the PDOS of different elements as indicate by the gray solid lines.

From the perspective of Mulliken charges, that of Li does not change significantly even with an increase in Li. The Mulliken charge of Li is +0.417e and +0.436e in $Li_{0.042}ZrNCl$ and $Li_{0.500}ZrNCl$, respectively. The Mulliken charge of Cl, which is in contact with Li, changes from -0.312e before insertion to -0.434e in $Li_{0.500}ZrNCl$, and thus Cl tends to acquire a more negative charge. In addition, not only Cl but also Zr undergo changes in Mulliken charge. In ZrNCl, Zr has +1.032e, while it decreases to +0.892e in $Li_{0.500}ZrNCl$, acquiring a negative charge approximately equivalent to that received by Cl.

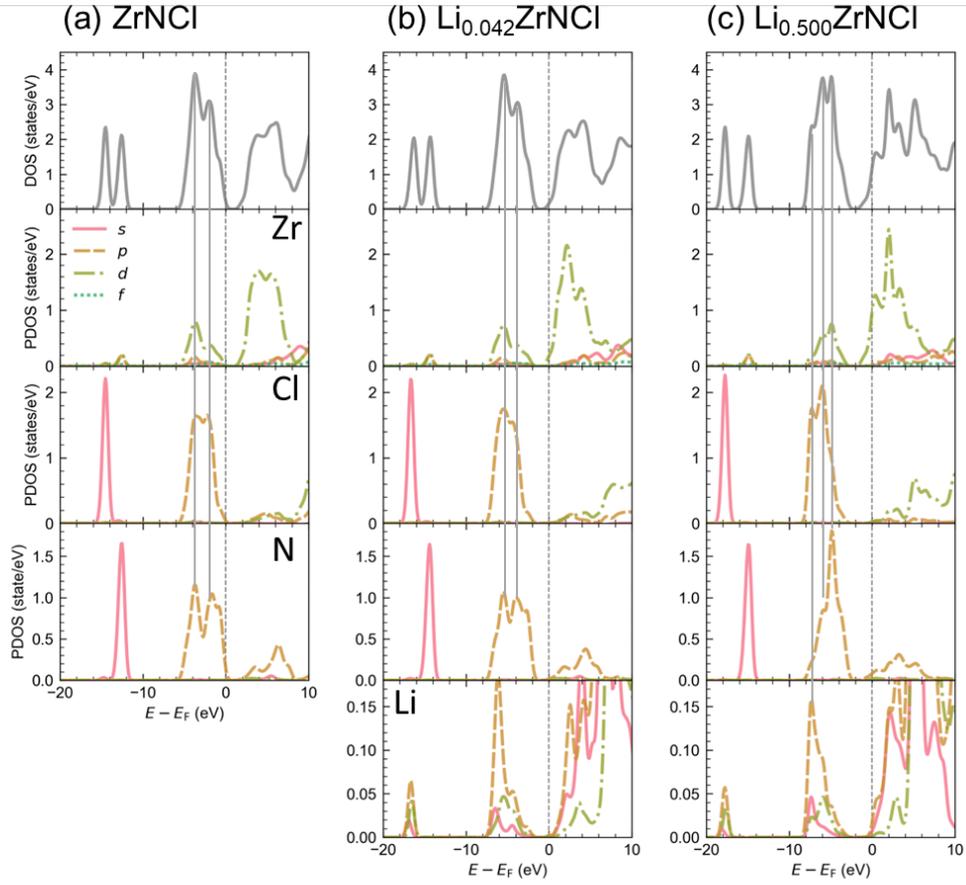

**Figure 9** Calculated total DOS and atom-PDOS for Zr, Cl, N, and Li for (a) ZrNCl, (b) Li$_{0.042}$ZrNCl, and (c) Li$_{0.500}$ZrNCl. The vertical dashed line corresponds to the position of Fermi energy ($E_F$). The total DOS has been scaled to be per ZrNCl.

The results of COOP analysis are shown in Fig. 10. The first row represents the total DOS, and the second, third, and fourth rows represents the COOP diagram for Zr-Cl, Zr-N, and Li-Cl interactions, respectively. In the case of Zr-N bonds, two types of bonding should be considered: bonds in the *ab*-plane and bonds in the *c*-direction. However, because they showed similar trends of changes, here we will only focus on Zr-N bonds in the *c*-direction. From Fig. 10, it can be observed that there is a bonding interaction between Cl atom and intercalated Li atom (forth row of Figs.10 (b) and (c)).

However, the COOP values between Li and Cl around -9~-7eV decrease with an increasing Li content. As a result, the BOP values slightly decrease from +0.082e in Li$_{0.042}$ZrNCl to +0.077e in Li$_{0.500}$ZrNCl. This is likely be attributed to the high electronegativity of Cl, suggesting that the Li-Cl bonding becomes more ionic by increasing the Li concentration.

As for the COOP between Zr and Cl, it is confirmed that bonding interactions between Zr and Cl decreases, and anti-bonding interactions occur with Li insertion. Consequently, the BOP values between Zr and Cl exhibit a monotonic decrease from +0.130e in ZrNCl, +0.099e in Li$_{0.042}$ZrNCl, and further to +0.058e in Li$_{0.500}$ZrNCl. In contrast, there is significant enhancement of the Zr-N bonding interaction at the energy denoted by "p1", where the Zr-Cl anti-bonding interactions are also strengthened. This increase of the Zr-Cl anti-bonding interaction would influence to the intercalation energies. The intercalation energy increases from -2.893 eV in Li$_{0.042}$ZrNCl to -2.218 eV in Li$_{0.500}$ZrNCl, which is thought to be correlated with the destabilization of the Zr-Cl bonding.

Based on the aforementioned discussion, the way to accept the Li intercalation by changing the electronic structure of ZrNCl is summarized as follows: Upon the intercalation of Li in ZrNCl, Cl and Zr accept electrons from Li, weakening the covalent bond between Cl and Zr. At the same time, Zr and N are forming a new chemical bonding state. Namely, the elements constituting the layered compounds synchronously change their chemical bonding and accept the Li intercalation, which was first revealed by the COOP analysis. The energy for intercalation increased with the increase in Li content, which is possibly correlated with the destabilization of the bonding between the Zr and Cl from the perspective of covalent bonding.

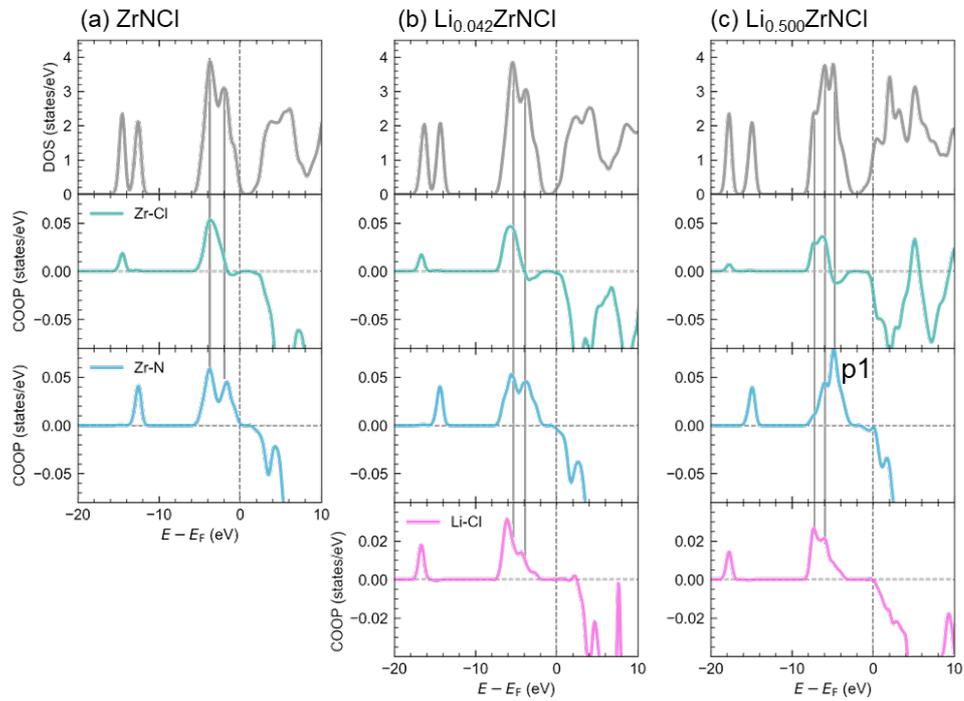

**Figure 10** Calculated total DOS and COOP diagram about Zr-Cl interactions, Zr-N interactions, and Li-Cl interactions for (a) ZrNCl, (b) Li$_{0.042}$ZrNCl, and (c) Li$_{0.500}$ZrNCl. Vertical dashed lines correspond to the position of Fermi energy ($E_F$). The total DOS has been scaled to be per ZrNCl.

## 4. Conclusion

In this work, we developed a code to perform COOP analysis based on all-electron DFT calculation with numeric atom-centered orbitals using FHI-aims. Then, we conducted the COOP analysis on molecular and solid-state systems: $F_2$, Si, and $CaC_6$, confirming the validity of the present method and demonstrating the chemical-bonding interpretation. Furthermore, we conducted the COOP analysis to investigate the changes in the chemical-bonding associated with a Li intercalation in three representative layered materials: graphite, $MoS_2$, and ZrNCl.

In graphite, it was clarified that Li intercalation leads to bonding interactions between Li and C, while C-C anti-bonding interactions emerge, leading to the destabilization of the host structure. As for 2H-$MoS_2$, while there was charge transfer from Li to S, the change in the chemical-bonding state was small. This small chemical-bonding change was consistent to the small change in the intercalation energy of Li to 2H-$MoS_2$. In ZrNCl, electronic states of all elements were affected by the Li intercalation, weakening the covalent bond between Cl and Zr. On the other hand, it was revealed that new bonding orbitals are formed between Zr and N, effectively capturing the effects of Li intercalation.

In conclusion, the COOP analysis enabled us to accurately understand the change in the characteristics of chemical-bonding associated with Li intercalation and elucidate the origin of stability in Li intercalated materials from a chemical bonding perspective.

**Acknowledgement**

This work was supported by MEXT/JSPS, KAKENHI Grant Number JP19H05787 and JST, CREST Grand Number JP-MJCR1993. Mr. Naoto Kawaguchi, The University of Tokyo, is acknowledged for fruitful discussion on the chemical-bonding of CaC$_6$, and Prof. Yoshihiro Iwasa, The University of Tokyo, is acknowledged for the helpful suggestions on the Li intercalation to ZrNCl.